\documentclass{tMPH2e} %
\usepackage[utf8]{inputenc}%
\usepackage{graphicx}%
\usepackage[usenames,dvipsnames]{color}
\newlength{\onefig}
\newcommand{\PageHeaders}{\markboth{A. Ninarello {\rm et al.}}{Molecular Physics}}%

\newcommand{\epsi}{\varepsilon}

\begin{document}%

\PageHeaders \articletype{SPECIAL ISSUE PAPER}%

\title{Structure and dynamics of coupled viscous liquids}%
\author{Andrea Ninarello, Ludovic Berthier, 
Daniele Coslovich\\ \vspace{6pt}%
Laboratoire Charles Coulomb UMR 5221, 
Universit{\'e} de Montpellier and CNRS, Montpellier, France \\
\received{``received''}}%

\maketitle

\begin{abstract}
We perform Monte-Carlo simulations to analyse the structure 
and microscopic dynamics of a viscous Lennard-Jones liquid coupled to a 
quenched reference configuration of the same liquid. 
The coupling between the two 
replicas is introduced via a field $\epsi$ conjugate 
to the overlap $Q$ between the two particle configurations. This allows us 
to study the evolution of various static and dynamic correlation functions 
across the $(\epsi, T)$ equilibrium phase diagram. 
As the temperature is decreased, we identify 
increasingly marked precursors of a first-order phase transition between 
a low-$Q$ and a high-$Q$ phase induced by the field $\epsi$. 
We show in particular that 
both static and dynamic susceptibilities have a maximum at a 
temperature-dependent value of the coupling field, which defines 
a `Widom line'.  
We also show that, in the high-overlap regime, 
diffusion and structural relaxation are strongly decoupled because
single particle motion mostly occurs via discrete hopping on the sites defined 
by the reference configuration. These results, obtained using conventional
numerical tools, provide encouraging signs that an equilibrium phase 
transition exists in coupled viscous liquids, but also demonstrate
that important numerical challenges must be overcome 
to obtain more conclusive numerical evidence.
\end{abstract}


\section{Introduction}
\label{sec:intro}

`Cloning' or `replicating' configurations in glassy systems has a 
long history, which parallels the use of replica calculations 
to deal with systems defined by interactions containing quenched
disorder~\cite{Mezard_1986}, such as spin glasses or 
disordered ferromagnets~\cite{Mezard_1986,Young_1998}. 
The relevance of replicas in the context of the glass transition 
of supercooled liquids has emerged more recently, and it originates from the 
proposal that disordered spin glass models and supercooled liquids 
belong to a unique universality class and undergo 
at low temperature a random first order transition (RFOT)
between a liquid phase and an ideal glass 
phase~\cite{thirumalai__1989}. Whereas 
this identification holds rigorously at the mean-field 
level~\cite{Charbonneau_Kurchan_Parisi_Urbani_Zamponi_2014}, 
much remains to be understood regarding the applicability of these
concepts in finite dimensions, where a variety of fluctuation effects could 
drastically modify the mean-field picture~\cite{Berthier_Biroli_2011}.  

In recent years, it has been realized that the relevance of 
the RFOT construction can be assessed using `extended' 
phase diagrams, where an additional control parameter is introduced 
to probe the RFOT physics in temperature regimes that are more easily
accessible to simulations and experiments. The general idea 
is that precursors of the putative thermodynamic glass transition
develop as the temperature is lowered in a bulk liquid, the 
glass phase being `metastable' with respect to the liquid phase
\cite{franz_phase_1997}.  
Therefore, the addition of well-chosen external fields could 
potentially stabilize this phase at temperatures well above the 
equilibrium glass transition and induce equilibrium phase transitions
which are direct byproducts of the RFOT physics.
In other words, external constraints might reveal 
growing structural correlations characterizing 
viscous liquids approaching the glass transition in a way that is not 
accessible to the standard tools used for 
simple liquids~\cite{hansen_theory_1986}. Studying constrained 
systems thus might be useful to reveal information about bulk 
systems, and constrained phase transitions may exist even 
in systems where no finite temperature Kauzmann transition 
exists \cite{zamponi}, the two issues being logically independent.   
Among the possible choices for such external constraints, 
the idea of pinning the position of a set of particles 
has received considerable attention 
recently~\cite{Bouchaud-JCP2004,montanari_rigorous_2006}, 
in a variety of geometries~\cite{berthier_static_2012} 
from finite cavities~\cite{biroli_thermodynamic_2008,cavagna_dynamic_2012,Hocky-PRL2012},
to fully random pinning~\cite{cammarota_ideal_2012,jack2012,Charbonneau_Charbonneau_Tarjus_2012,kob_probing_2013,long_giulio,jack2014,kob:2015}
or amorphous walls~\cite{Kob-NatPhys2011,Hocky_Berthier_Kob_Reichman_2014}.  

In this article, 
we analyse the case where an external coupling to a quenched reference 
configuration is introduced via a field $\epsi$ conjugate to the 
overlap $Q$ between the two copies of the system. 
This situation has been studied  
analytically in mean-field models~\cite{kurchan93,franz_phase_1997,long_giulio,FP2013,tarzia}
and finite dimensional liquids~\cite{cardenas_constrained_1999}, 
and it was also studied in computer 
simulations~\cite{franz_parisi1998,cardenas_constrained_1999,cammarota_phase-separation_2010,berthier_overlap_2013,berthier2015}. 
The `annealed' situation where the coupling is 
between two evolving copies of the system has also received 
attention from a number of groups 
\cite{mezard99,FP2013,bomont_probing_2014,bomont_investigation_2014,berthier_overlap_2013,parisi_liquid-glass_2014,garrahan2014}. 
For the quenched coupling of interest in this work, 
the phase diagram in the $(\epsi, T)$ plane has been 
established in the mean-field approximation~\cite{franz_phase_1997}.
In that case, an ordinary first-order transition line emerges from 
the bulk glass transition temperature, which separates a 
low-overlap (uncorrelated) phase from a large-overlap (localized)
phase. This first-order transition line ends at a second-order critical 
point which, if present, should be in the same universality class as the 
random field Ising model~\cite{tarzia,FP2013}.    
 
Our central goal is to use standard computational tools 
to explore the high-temperature region of the $(\epsi, T)$ 
phase diagram where conventional Monte-Carlo simulations are sufficient
to achieve thermalization of the coupled system. By analysing 
several static and dynamic correlation functions, we find 
a number of distinctive precursors of the first-order transition line, 
which allow us to define a `Widom line' for our system. We also
find that the microscopic dynamics in the high-overlap regime 
is markedly different from the one of bulk supercooled liquids, 
because single particle motion occurs mostly via spatially uncorrelated, 
discrete hopping on the sites defined by the reference configuration. 
Because this diffusion process is very slow, we conclude that an equilibrium 
exploration of the two phases of the model at lower temperatures 
is impossible using conventional means.  

Our paper is organised as follows. In Sec.~\ref{sec:methods} we briefly 
present the model and our numerical methods.
In Sec.~\ref{sec:statics} we analyse the static properties
of the model, whereas Sec.~\ref{sec:dynamics} presents results
for dynamic correlation functions. 
Sec.~\ref{sec:conclusion} closes the paper with some perspectives for future 
work.

\section{Model and numerical methods}
\label{sec:methods}

We performed Monte-Carlo simulations~\cite{berthier_monte_2007}
of the Kob-Andersen glass-forming
model~\cite{kob_testing_1995}, which is a $80:20$ binary mixture of 
particles interacting via the Lennard-Jones potential 
\begin{equation}
v_{\alpha\beta}(r)=4\epsilon_{\alpha\beta}\left[\left( 
\frac{\sigma_{\alpha\beta}}{r}\right)^{12}-\left( \frac{\sigma_{\alpha\beta}}{r}
\right)^{6} \right].
\end{equation}
where $\alpha, \beta= \{A, B\}$ denote the particles' species.
The potential is shifted to ensure its continuity at the cut-off distance 
$r_c=2.5$.
In the following, all quantities will be expressed in reduced units, 
selecting $\epsilon_{AA}$ and $\sigma_{AA}$ as units of energy and 
distance, respectively, and the Boltzmann constant is $k_B=1$.
The interaction parameters are $\sigma_{AA}=1.0$, $\sigma_{AB}=0.8$, 
$\sigma_{BB}=0.88$, $\epsilon_{AA}=1.0$, $\epsilon_{AB}=1.5$ and 
$\epsilon_{BB}=0.5$. They have been chosen to 
prevent the mixture from crystallizing at high densities and 
low temperatures. We have studied this particular model because 
it is a well-studied and well-characterized glass-forming 
system~\cite{kob_testing_1995,berthier_monte_2007,entropy2014}.
Exploratory runs performed using molecular dynamics methods instead 
of Monte-Carlo suggest that the dynamical results that we report
do not depend qualitatively on the specific choice of a microscopic
dynamics. 

We studied a system of $N=1000$ particles in each copy. 
We work in three spatial dimensions using  
periodic boundary conditions at a number density $\rho= N/V = 1.2$,
where $V$ is the volume of the simulation box.
In our Monte-Carlo simulations, we perform sequential
single particle displacements by drawing random  
displacements within a cube of 
linear size $\Delta r_\textrm{max}$ centered on the particle's position. 
The proposed displacement
is accepted according to the Metropolis criterion.
We chose $\Delta r_\textrm{max}=0.12$, which maximizes the 
diffusion constant of the system in the unconstrained mixture at $T=1.0$,
which represents the onset temperature for the bulk Lennard-Jones 
system.
The time unit in our Monte-Carlo simulations is defined as $N$
attempts to move a particle.  

To study the coupled system, we first define the overlap $Q_{12}$
between two configurations, 1 and 2, as
\begin{equation}
  \label{eq:Q}
  Q_{12} = \frac{1}{N} \sum_{i,j} \theta(a - | {\bf r}_{1,i} - 
{\bf r}_{2,i}|),
\end{equation}
where $\theta(x)$ is the Heavyside function, $a=0.3$ is a coarse-graining 
distance for comparing the density profiles, and ${\bf r}_{\alpha,i}$ 
denotes the position of particle $i$ in the copy $\alpha \in \{1,2\}$. 

Our numerical procedure to study the coupled glassy problem 
is to draw a series of equilibrium configurations 
$\{ {\bf r}_1 \}$ at temperature $T_1$, which serve as reference
configurations. We then study 
a second copy of the system, $\{ {\bf r}_2 \}$, which evolves 
at temperature $T_2$ and
is coupled to a given reference configuration via a 
field $\epsi$ conjugate to the overlap between the two 
configurations. Only the second copy is allowed to evolve in this step.
The total Hamiltonian thus reads 
\begin{equation}
  \label{eq:H}
 H_{ \{ {\bf r}_1 \}} ( \{ {\bf r}_2 \} ) =
H_{\rm LJ} (\{ {\bf r}_2 \})  - \epsi Q_{12},
\end{equation}
where
\begin{equation}
H_{\rm LJ}( \{ {\bf r} \}  ) = \frac{1}{2} \sum_{\alpha,\beta} \sum_{i=1}^{N_\alpha}
\sum_{j=1}^{N_\beta} v_{\alpha \beta}({\bf r}_{ij})
\end{equation}
denotes the Lennard-Jones Hamiltonian for a single 
copy $\{ {\bf r} \}$, and $\epsi>0$ biases the coupled system 
towards higher values of the overlap. 
In our study, we use equal temperatures for the quenched reference 
configurations and the system under study, $T_1 = T_2 = T$,
although other choices are possible~\cite{franz_phase_1997,berthier2015}.
The Hamiltonian (\ref{eq:H}) contains quenched disorder, because the 
positions of all particles in copy 1, $\{ {\bf r}_1 \}$, are held fixed. 
Therefore, after thermal average is performed at temperature $T$ for a given 
realisation of the disorder, we need to perform an average over 
independent configurations of the quenched copy.
We found that the disorder average plays a negligible role for the system 
size and the temperature range studied in this work.
Therefore, we only needed to average over a small number of independent 
quenched configurations to obtain accurate static and dynamic properties of 
the system. A much more demanding disorder averaging procedure would
be needed at lower temperatures~\cite{berthier_overlap_2013,berthier2015}. 

Finally we find that thermalising the constrained 
system defined by the Hamiltonian (\ref{eq:H}) becomes 
increasingly difficult when $\epsi$ is large and/or temperature 
is low. For a given state point defined by $(\epsi,T)$, 
we consider that a system is equilibrated if the two following criteria
are met. First, we require that the average of the overlap 
evaluated over a time window of the order of 
$10^{5}$, displays no systematic drift. Second, we require that 
particles move an average distance of about 3 particle diameters. 
These empirical criteria are chosen so that 
both single particle dynamics and overlap fluctuations 
are accurately sampled in our simulations. Remark that 
even when the overlap value is large and the system
explores a limited part of the configurational part close
to the reference configuration, it is important to 
perform very long simulations to ensure that both static and dynamic
quantities are probed accurately. 
As a consequence, we have not been able to study
meaningful parts of the $(\epsi,T)$ phase diagram 
for temperatures below $T = 0.7$ (our data at $T=0.6$ barely satisfy
our criteria). Note that this temperature only represents 
a modest degree of supercooling for the unconstrained system and is far above
the mode-coupling temperature $T_{\rm mct} \approx 0.435$, but 
this already represents a challenging situation for the coupled system. 

\section{Static properties}
\label{sec:statics}

\begin{figure}\centering
\includegraphics[width=7.7cm,clip]{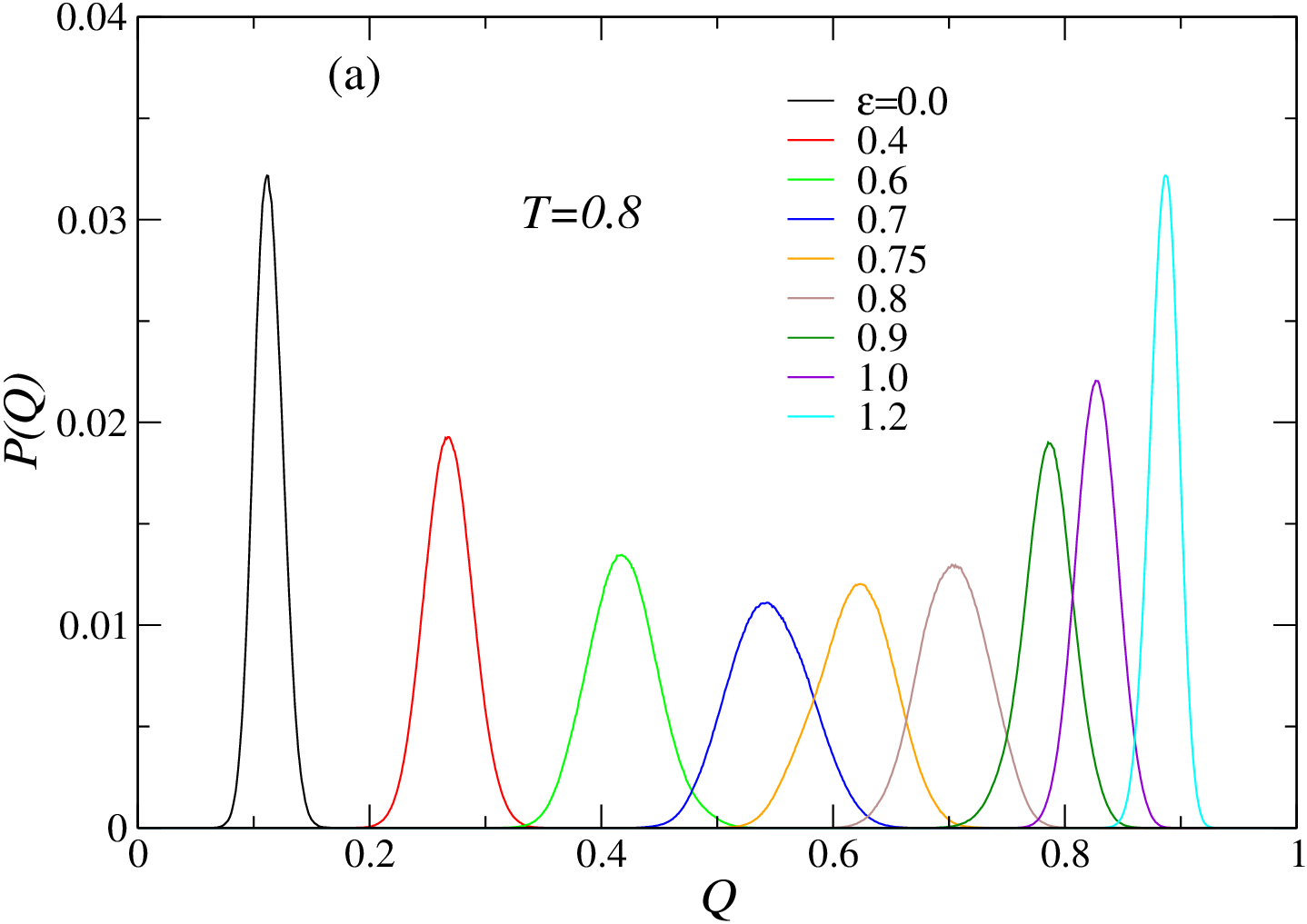}
\includegraphics[width=8cm,clip]{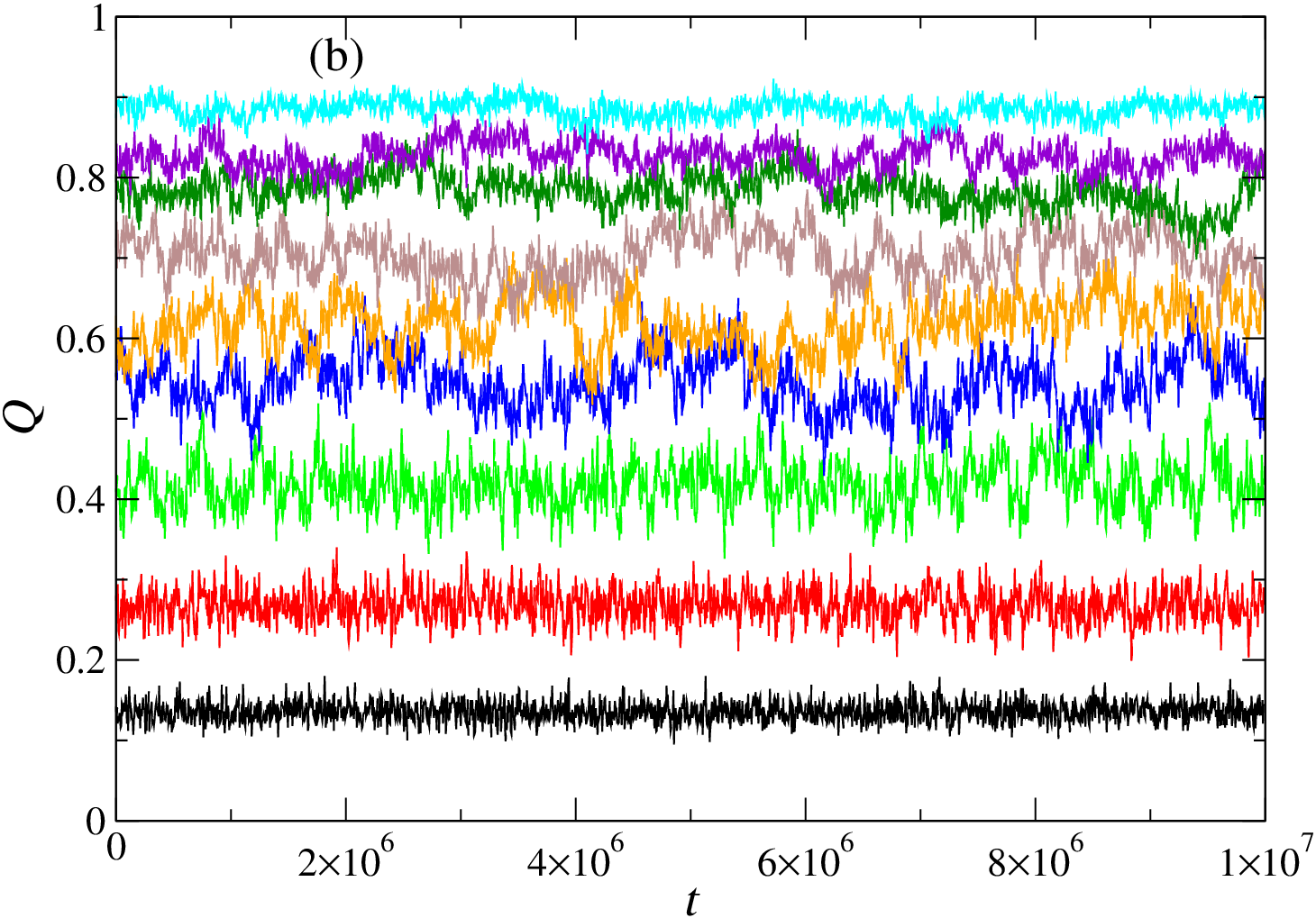}
\caption{\label{fig:Pq} (a) Probability distribution of the overlap 
$P(Q)$ at several $\varepsilon$ values along 
the isotherm $T=0.8$. (b) Typical time series of 
the overlap $Q(t)$ for the same parameters (the bulk 
alpha-relaxation time at $T=0.8$ is about $\tau \approx 2 \cdot 10^3$). 
The overlap fluctuations 
are broader and slower near $\epsilon^* \approx 0.7$.}
\end{figure}

We start by studying the static properties of the system defined 
by Hamiltonian (\ref{eq:H}), varying $\epsi$ and $T$. 
Intuitively, we expect that large values of $\varepsilon$ will localize 
the particles close to the sites defined by the position of the 
particles in the quenched reference configuration in order to 
increase the overlap $Q$ between the two replicas.
In Fig.~\ref{fig:Pq}(a), we show the probability distribution of the 
overlap, $P(Q)$, along the representative isotherm $T=0.8$
for increasing values of the coupling field $\epsi$. 
This temperature is slightly below the onset temperature ($T\approx1.0$) 
of slow dynamics in the 
unconstrained system with $\varepsilon=0$.
As the coupling $\varepsilon$ increases, we find indeed that $P(Q)$ 
becomes centered around increasingly larger values of $Q$.
As discussed in Sec.~\ref{sec:intro}, we may expect that 
the system undergoes a 
first-order phase transition between a low-$Q$ phase and a high-$Q$ 
phase when temperature is low enough. Clearly, the data 
in Fig.~\ref{fig:Pq}(a) show that the overlap increases smoothly 
with $\epsi$ at this temperature and the phase transition,
if present, must occur at lower temperatures.  

However, we notice that the distributions are narrow for both 
small and large values of $\epsi$ but broaden considerably 
at intermediate values corresponding to 
intermediate overlap values, $Q \approx 0.5$.   
Such a broadening of the distributions for intermediate values of 
$\varepsilon$ is consistent with the idea that the system 
approaches the critical temperature from above.

A confirmation of this intuition is shown in 
Fig.~\ref{fig:Pq}(b) which illustrates the typical time evolution of the order 
parameter $Q(t)$ in the course of the Monte-Carlo simulations.
We observe that the overlap displays slow fluctuations of large amplitude 
for intermediate values of $\varepsilon$, 
whereas temporal correlations and large excursions are strongly suppressed 
when the replicas are strongly coupled in the high-$Q$ regime, 
or nearly uncorrelated at low $\epsi$. 
These qualitative observations suggest that the dynamics of the
coupled system display a non-trivial variation as a function 
of the coupling. The large fluctuations of the global overlap 
$Q$ represent a first source of dynamic slowing down for 
the coupled system, which we interpret as a form of critical slowing 
down, expected in the vicinity of a second-order critical point. 
Notice that the time series shown in Fig.~\ref{fig:Pq} cover a window
of about $10^7$ timesteps, whereas the bulk alpha-relaxation time 
of the unconstrained system at the same temperature is $\tau \approx 
2 \cdot 10^3$, which already demonstrates that accessing 
the properties of the coupled system in conventional numerical
simulation is challenging, even at temperatures far above the critical
point.    
 
\begin{figure}
\centering
\includegraphics[width=8cm,clip]{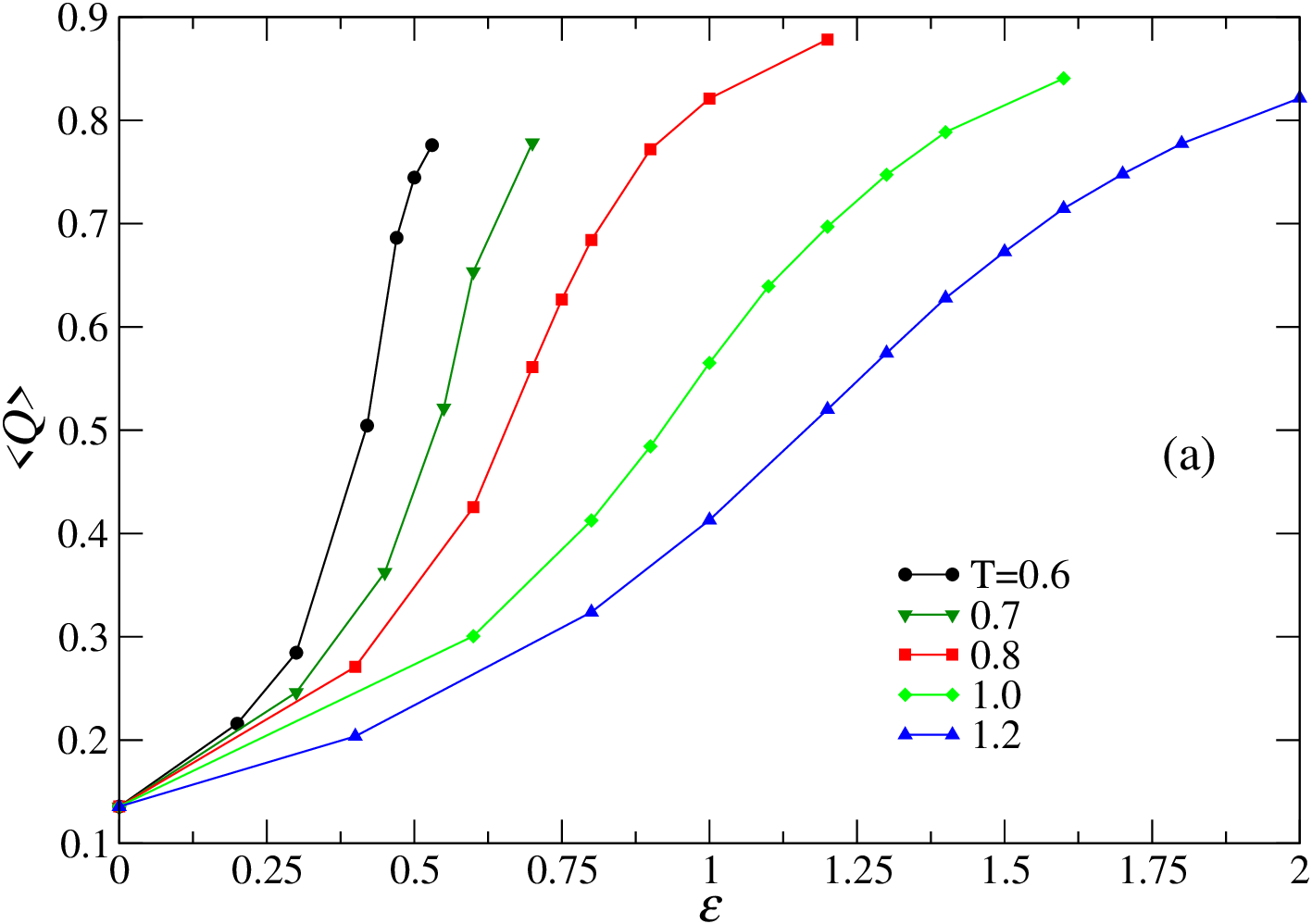}
\includegraphics[width=8.1cm,clip]{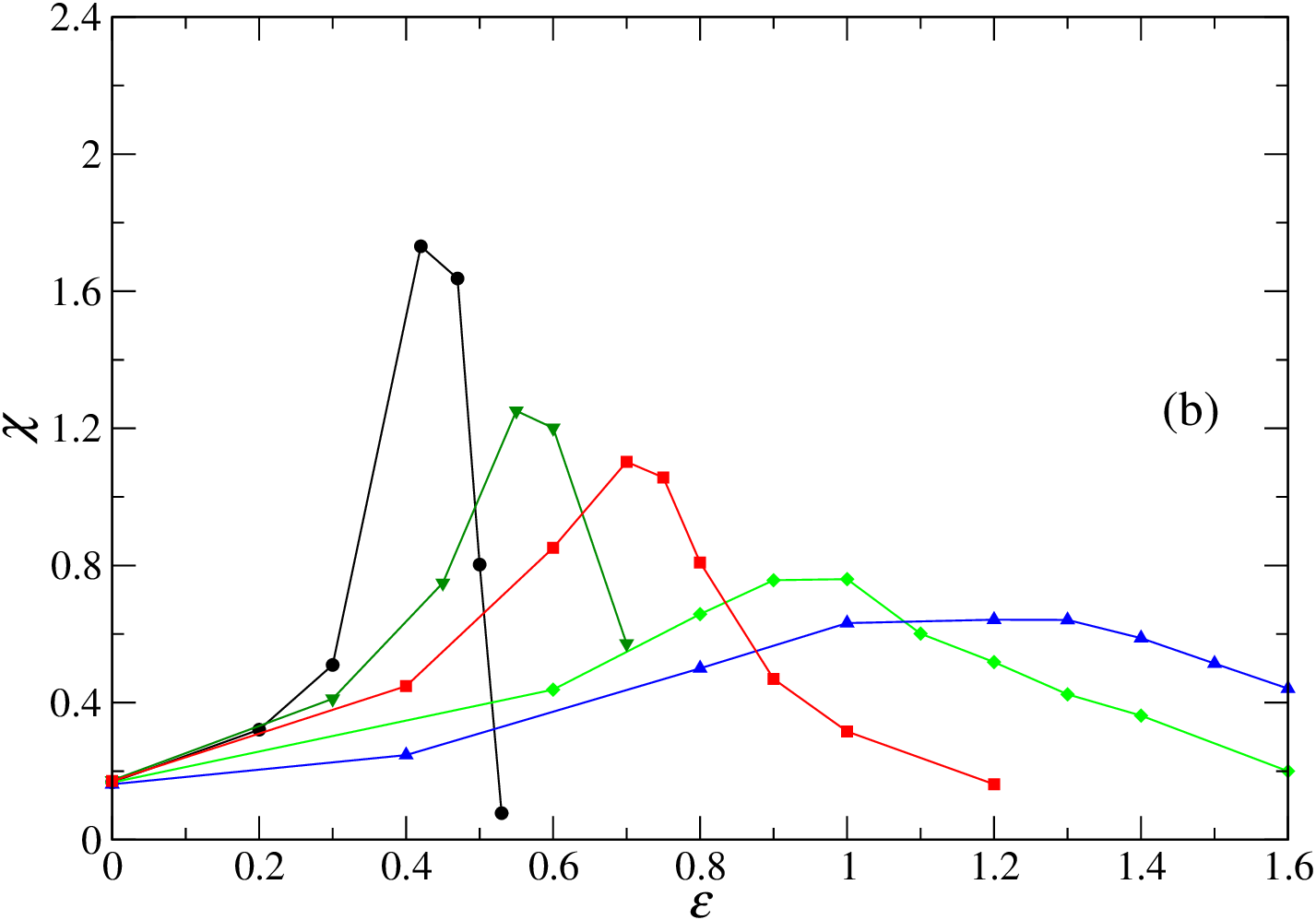}
\caption{\label{fig:qave} (a) Average overlap $\langle Q \rangle$ and 
(b) overlap susceptibility $\chi$ as a function of $\varepsilon$ for
various temperatures. The crossover from uncorrelated to localized 
regimes occurs at lower $\epsi$ when $T$ decreases, and becomes 
sharper.}
\end{figure}

In Fig.~\ref{fig:qave}(a) we show the average overlap $\langle Q\rangle$ 
as a function of $\varepsilon$ for all studied temperatures, 
where the moments of the distribution are defined as
\begin{equation}
\langle Q^n \rangle = \int_0^1 dQ P(Q) Q^n.  
\end{equation}
We find that the coupling needed to localize the system in the 
high-overlap regime decreases rapidly upon decreasing the temperature.
This directly implies that the thermodynamic driving force 
which allows the system to escape from a randomly chosen reference
configuration also decreases with the temperature. This observation
is at the core of the RFOT description of the glass 
transition~\cite{thirumalai__1989}. 
  
Moreover, we observe that 
the crossover from the low-$Q$ regime to the high-$Q$ one becomes 
sharper and better defined as $T$ decreases.
This is again consistent with the approach to a first-order phase transition 
separating a normal, low-$Q$ phase and a localized, high-$Q$ phase.
To identify this crossover more precisely, we evaluate the overlap 
susceptibility, 
\begin{equation}
\chi = N[\langle Q^2 \rangle - \langle Q \rangle^2].
\end{equation}
The results are shown in Fig.~\ref{fig:qave}(b).
For each studied temperature, $\chi$ displays a well-defined maximum at 
a coupling $\epsi^*=\epsi^*(T)$, which we identify as the location 
of the crossover.
The locus of $\epsi^*(T)$ in the $(\varepsilon,T)$ plane represents 
the equivalent of the Widom line~\cite{domb_phase_1972} for conventional
phase transitions, as it represents the continuation of the transition 
line above the critical point. It should be clear that our 
interpretations are consistent with the mean-field calculations, but
that our numerical results do not establish that an equilibrium
phase transitions actually exists for this system. An alternative view is 
that the clear crossover line that we detect remains 
a smooth crossover at any finite temperatures \cite{jack2012}. 
Deciding which is the correct scenario is beyond the scope of the present work.

\begin{figure}
\centering
\includegraphics[width=8cm,clip]{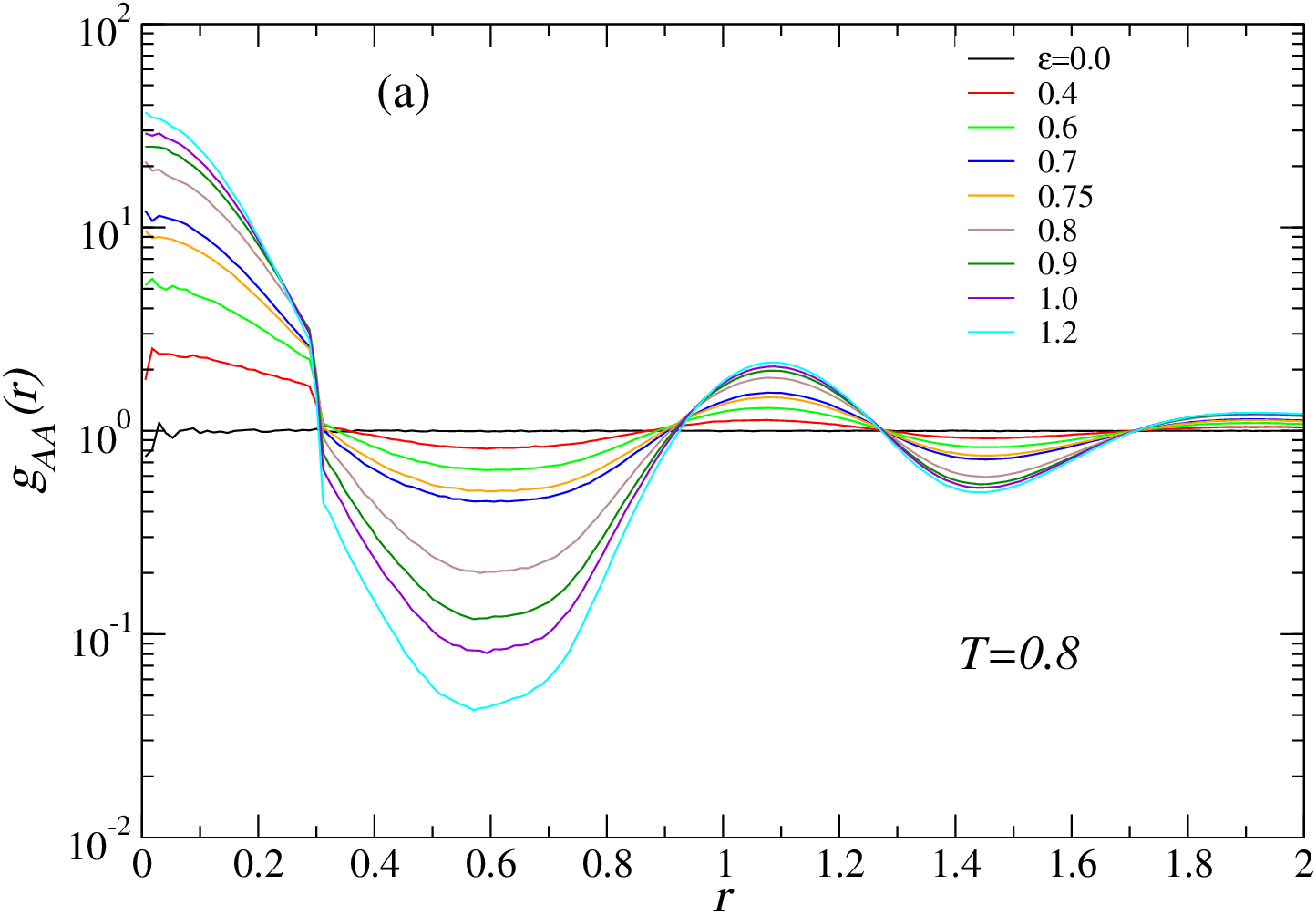}
\includegraphics[width=8cm,clip]{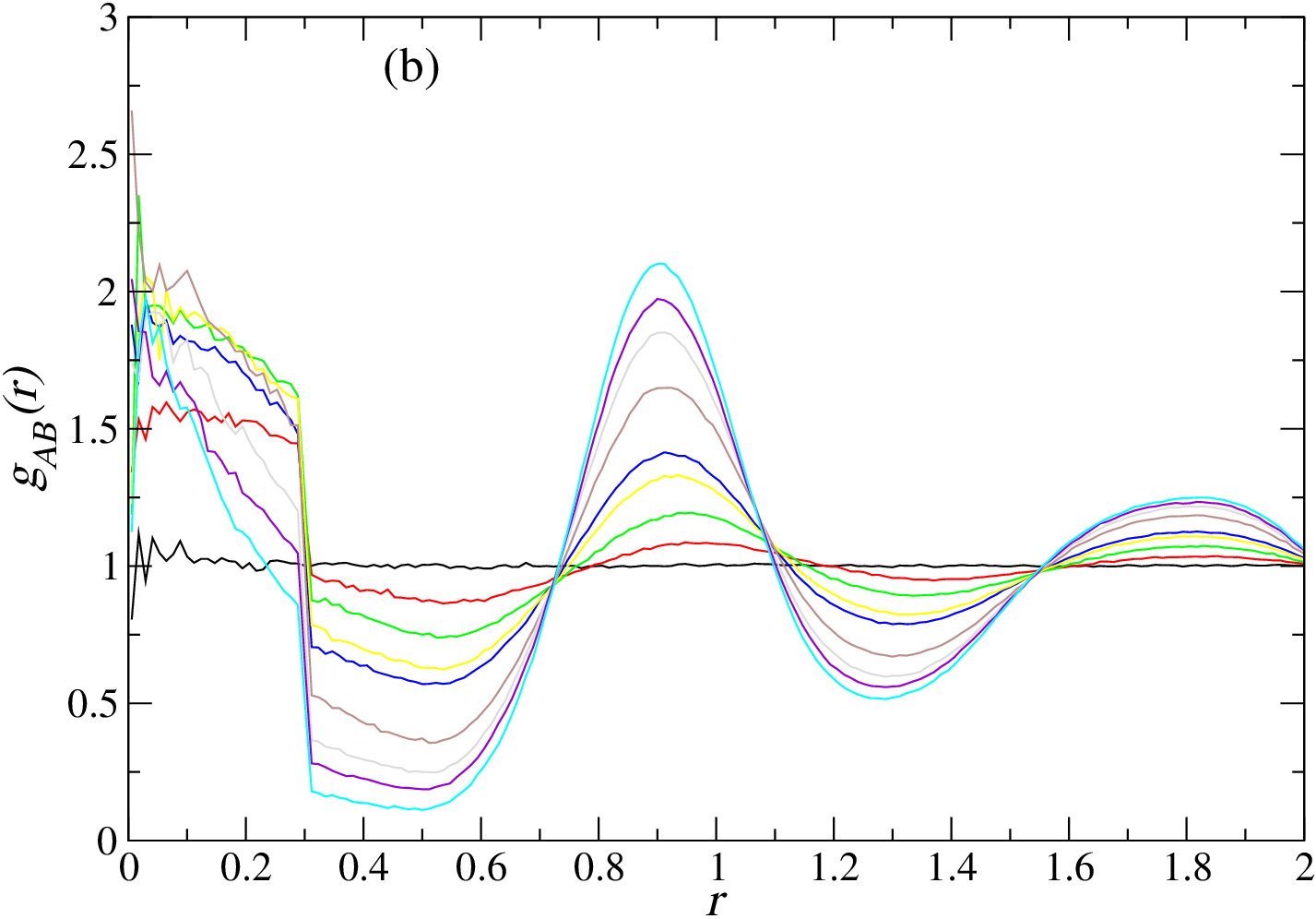}
\caption{\label{fig:gr} Radial distribution functions between quanched 
and fluid replicas $g_{AA}(r)$ and $g_{AB}(r)$, see 
Eq.~(\ref{eq:gr}). Both functions develop a clear maximum near 
$r=0$ as $\epsi$ increases, reflecting the global increase
of the overlap via the formation of `dimers'. Whereas the 
probability to form $AA$ dimers increases with $\epsi$, 
$AB$ dimers are more numerous near $\epsi^* \approx 0.7$, but 
are suppressed at larger coupling.}
\end{figure}

The coupling parameter $\varepsilon$ in the Hamiltonian, Eq.~\eqref{eq:H}, 
allows one to bias the system towards higher values of the overlap $Q$.
The small value $a=0.3$ used to evaluate the overlap leads to the 
formation of `dimers' formed when a particle in the liquid is localized 
close to a particle in the reference configuration. We note that 
higher order clusters (such as trimers) 
have an enormous energy cost in the studied range of 
temperatures and couplings, and do not form.
To investigate this point more quantitatively, we 
have evaluated the radial 
distribution functions between quenched and liquid replicas at
various $T$ and $\epsi$, 
\begin{equation}
g_{\alpha \beta}(r) = \frac{1}{V N_\alpha  N_\beta } \sum_{i=1}^{N_\alpha} 
\sum_{j=1}^{N_\beta} \delta (r - | {\bf r}_{1,j}  - {\bf r}_{2,i}  |  ).
\label{eq:gr} 
\end{equation}
In Fig.~\ref{fig:gr}(a) we show the radial distribution function between
the large particles in the two copies, while 
Fig.~\ref{fig:gr}(b) shows correlations between large particles 
in the liquid and small particles in the reference configuration.
As $\varepsilon$ increases, a well-defined peak develops near $r\approx 0$,
which quantifies the increasing localisation of particles 
near the sites defined by the reference configuration.
The global overlap $Q$ is in fact directly related to the area 
under these peaks, and the pair correlation functions 
shown in Fig.~\ref{fig:gr} are the natural outcomes
of integral equation approaches for the problem 
of coupled glassy systems \cite{cardenas_constrained_1999,bomont_probing_2014}. 
The pair correlation functions develop a clear minimum near 
$r\approx 0.6$, indicating the formation of tight dimers of liquid and 
quenched particles.
This phenomenon resembles the formation of clusters in suspensions of 
neutral~\cite{mladek_formation_2006} and 
charged~\cite{coslovich_ultrasoft_2011} ultrasoft colloidal particles.
Finally, we note that the kink at $r=a$ is a direct result 
of the Heavyside function used to couple the two 
replicas and would be absent if a 
smoother coupling (for instance a Gaussian function) had been used.

The comparison between $g_{AA}(r)$ and $g_{AB}(r)$ shows that 
the peaks at short distance in these functions behave somewhat 
differently. First, the overall amplitude of the two peaks 
is very different, 
as testified by the different vertical scales used to represent
these two functions in Fig.~\ref{fig:gr} 
(the amplitude at $r=0$ is about 10 times larger for $AA$ than for 
$AB$ pairs). This shows that particles in the liquid replica tend to localize
to a site of the reference configuration occupied by a particle 
of the same type when the coupling is strong, although the 
probability to form `defective' $AB$ dimers is not negligible.  
A more careful analysis of the peaks also shows that the peak 
amplitude in $g_{AA}$ increases monotonically with $\epsi$ whereas
the one in $g_{AB}$ is maximum near the crossover value 
$\epsi^* \approx 0.7$. This observation suggests that the large 
fluctuations of the overlap revealed in Fig.~\ref{fig:Pq} 
when $\epsi \approx \epsi^*$ likely correspond
to a large variety of localized configurations, where dimers 
of various kinds can be formed. When the coupling becomes stronger, 
the system resembles increasingly to the reference configuration  
and $AB$ dimers are suppressed, at the expense of $AA$ and $BB$ dimers.
Indeed, we find that $g_{BB}(r)$ essentially tracks the behaviour
of $g_{AA}(r)$. The conclusions are also supported by the decomposition of the
global overlap into contributions coming from the various species. 

\section{Microscopic dynamics}
\label{sec:dynamics}

\begin{figure}\centering
\includegraphics[width=8cm,clip]{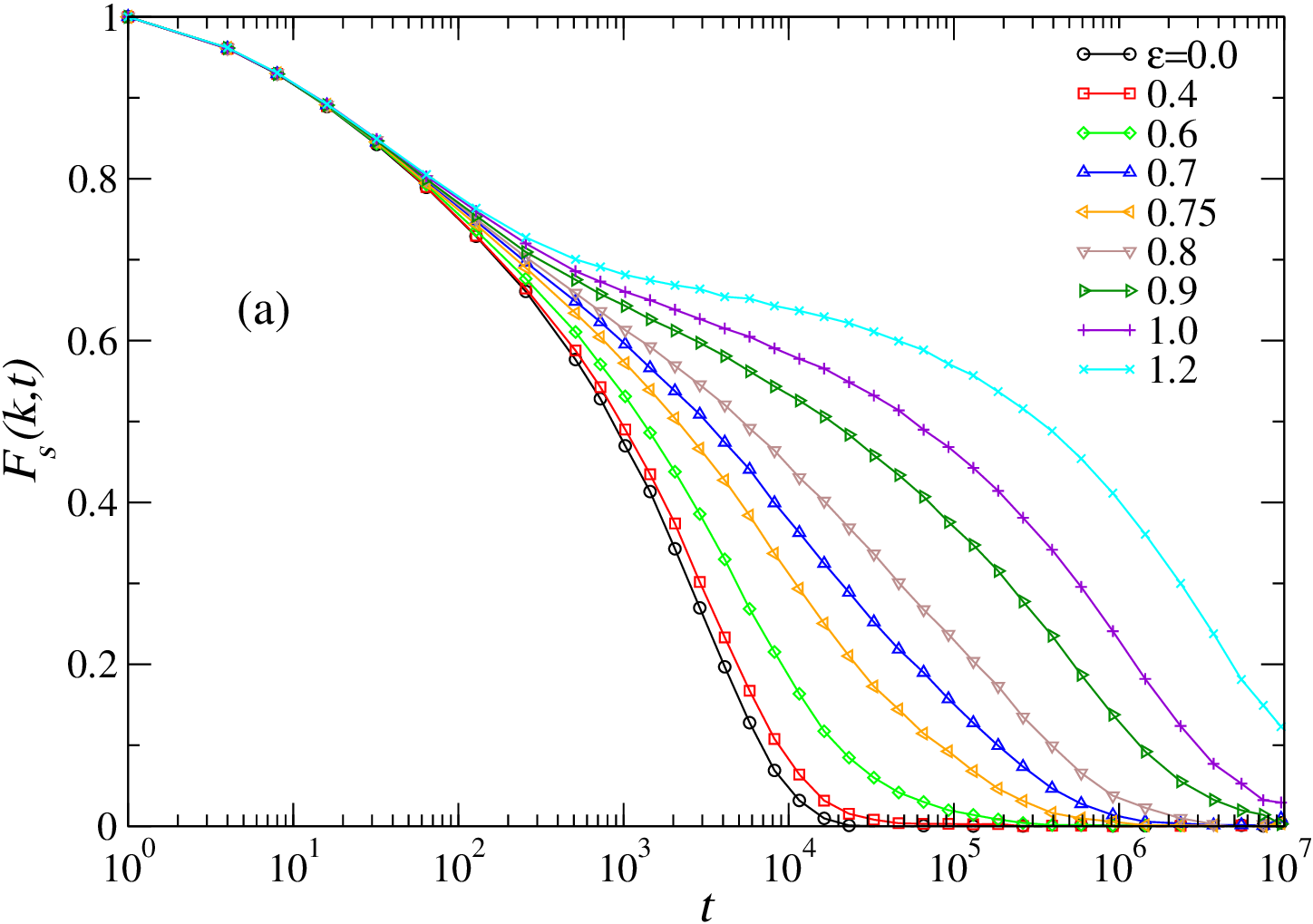}
\includegraphics[width=8cm,clip]{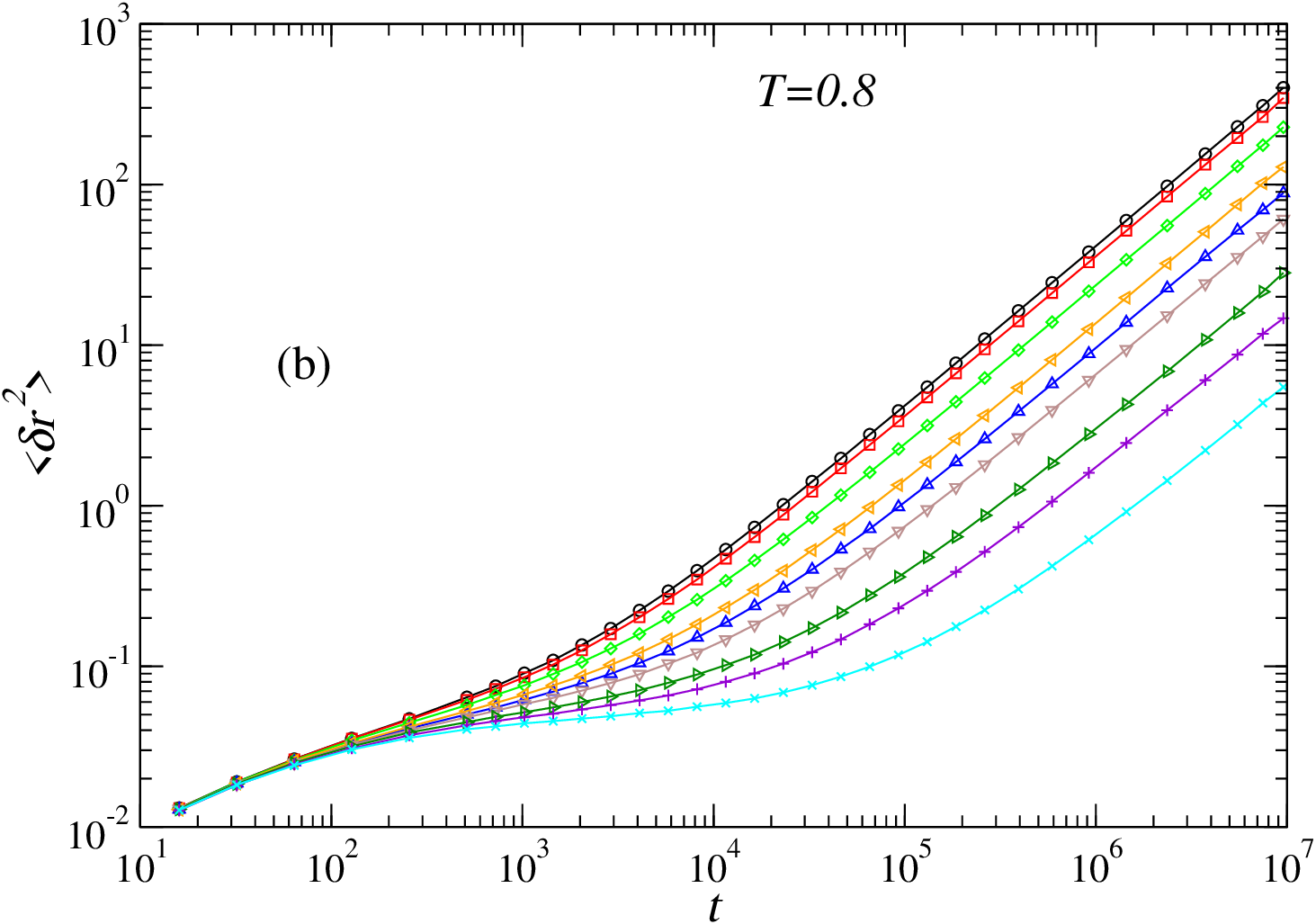}
\caption{\label{fig:fskt_msd} (a) Self-part of the intermediate 
scattering function $F_s(k,t)$ evaluated at $k=7.4$ along the isotherm
$T=0.8$ and various $\epsi$ coupling values.
(b) Mean-squared displacement $\delta r^2(t)$ for the same state points.
The single particle dynamics slows down monotonically by increasing 
$\epsi$, revealing the increasing localisation of the particles.}
\end{figure}

Close inspection of the time series of the overlap in Fig.~\ref{fig:Pq}(b) 
suggests that the dynamics of the liquid replica becomes increasingly 
sluggish upon increasing the coupling.
In the previous section, we (tentatively) attributed this slowing down
to the crossing of a Widom line where overlap fluctuations are very 
broad. However, we also noticed that dimers of like-particles form 
with increasing coupling, which corresponds to an 
increasing localization of the liquid particles on the quenched sites defined 
by the reference configuration. In this section, we will show 
that this localisation dramatically impacts
the available relaxation pathways.

To characterize the single-particle dynamics of the liquid replica, 
we evaluate the self-part of the intermediate scattering function 
\begin{equation}
F_s(k,t)= \frac{1}{N} \left\langle  \sum_i \exp[i 
{\bf k}\cdot ({\bf r}_i(t)-
{\bf r}_i(0)]\right\rangle,
\end{equation}
and the mean-squared displacement 
\begin{equation}
\delta r^2(t) = \frac{1}{N} \left\langle 
\sum_i |{\bf r}_i(t) - {\bf r}_i(0)|^2 \right\rangle.
\end{equation}
Both time correlation functions are evaluated for the large 
$A$ particles (the majority species) along the 
representative isotherm $T=0.8$. The results are displayed in 
Fig.~\ref{fig:fskt_msd}.

\begin{figure}\centering
\includegraphics[width=8cm,clip]{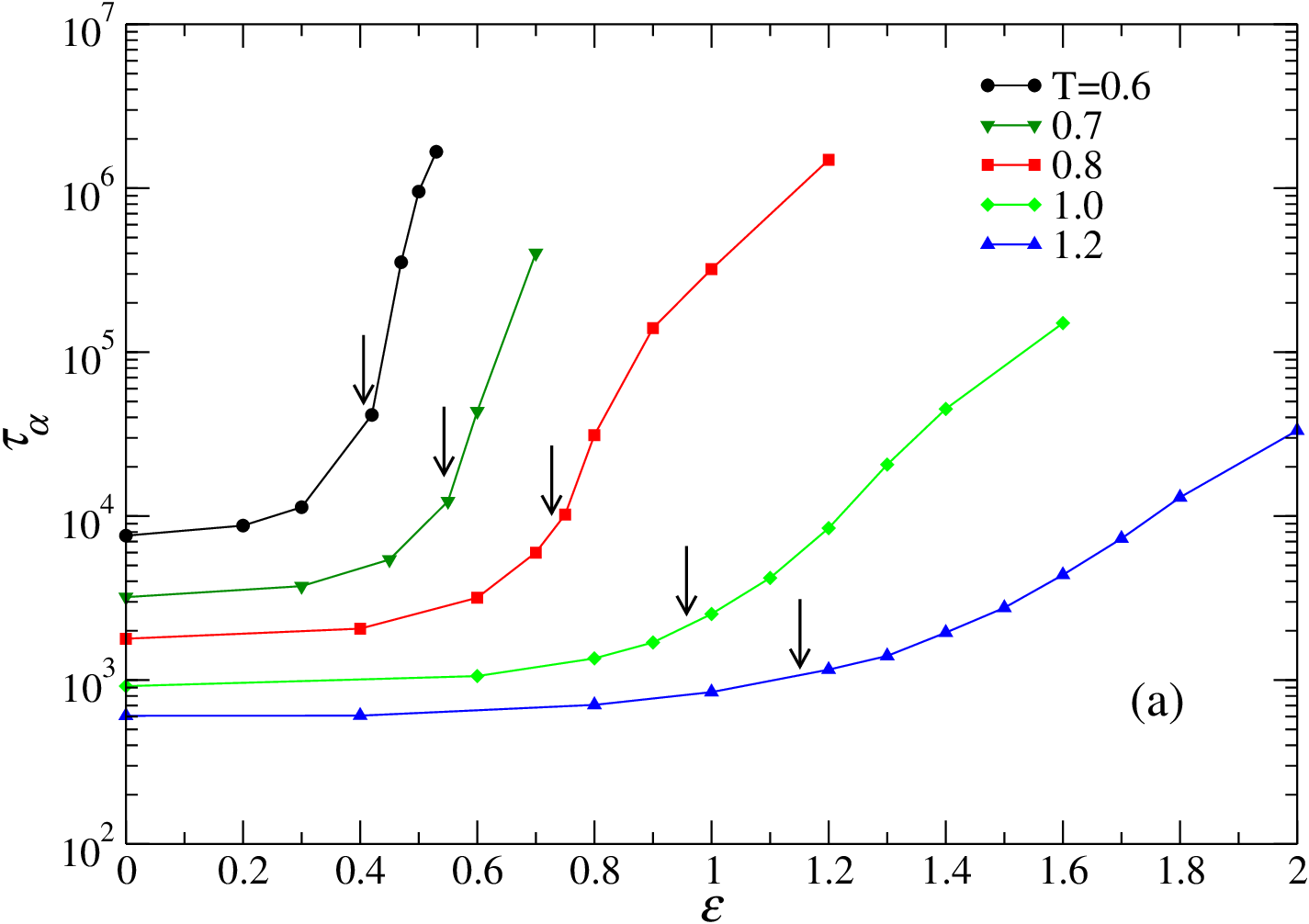}
\includegraphics[width=8cm,clip]{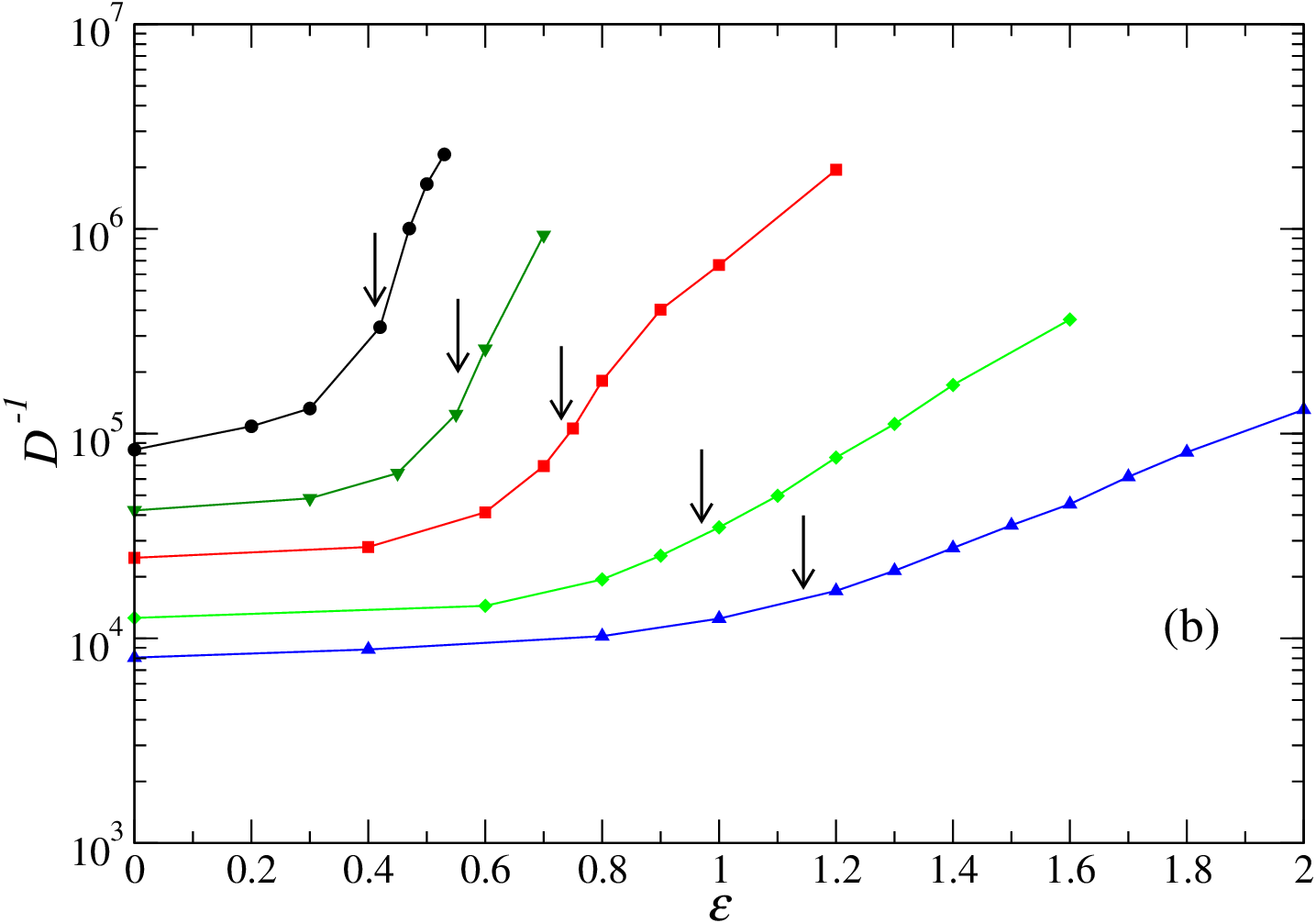}
\caption{\label{fig:tau_D} Evolution of the (a) 
structural relaxation times $\tau_\alpha$  
and of (b) the diffusion coefficients $D$ 
for large particles as a function of $\varepsilon$ for
different temperatures. The dynamics slows down dramatically 
when the crossover $\epsi^*(T)$ (shown with arrows)
is crossed. Note that the change 
in $\tau_\alpha$ and $D^{-1}$ is quantitatively different, as 
demonstrated in Fig.~\ref{fig:se_se}.}
\end{figure}

The unconstrained system displays practically no caging at this 
temperature, in agreement with the results of 
Ref.~\cite{berthier_monte_2007} on the Monte-Carlo dynamics of the 
same model. 
Upon increasing $\varepsilon$, however, the dynamics slows down significantly.
The relaxation of $F_s(k,t)$ at long times becomes increasingly 
stretched and eventually develops a clear plateau at intermediate 
times for the highest values of $\varepsilon$.
A similar plateau is observed in the mean-squared displacement $\delta r^2(t)$.
A comparison of the mean squared displacement for $A$ and $B$ 
particles (not shown here) reveals that in the high-$Q$ regime,
the smaller $B$ particles are somewhat less strongly localized 
on the quenched sites and can flow more easily than the large particles. 
The distinct dynamics between large and small particles is also 
observed in the bulk system at low temperatures \cite{kob_testing_1995}. 

As usual, we define the structural relaxation times $\tau_\alpha$ from the 
condition $F_s(k,\tau_\alpha)=1/e$ and the diffusion coefficients $D$ from 
the Einstein relation $\lim_{t \to \infty} \delta r^2(r) = 6 D t$.
In Fig.~\ref{fig:tau_D} we show the $\varepsilon$-dependence of $\tau_\alpha$ 
and $D^{-1}$ for $A$ particles.
For each studied temperature, we observe a marked increase of both 
$\tau_\alpha$ and $D^{-1}$ as the coupling approaches the crossover value 
$\varepsilon^*$, determined from the analysis of the overlap 
susceptibility (see Sec.~\ref{sec:statics}).
Upon further increasing $\varepsilon$, both quantities continue to 
increase steadily, and the single particle dynamics slows down 
dramatically. 
Note that all data points in Fig.~\ref{fig:tau_D} 
are fully equilibrated  according 
to the criteria detailed in Sec.~\ref{sec:methods}.

\begin{figure}\centering
\includegraphics[width=8cm,clip]{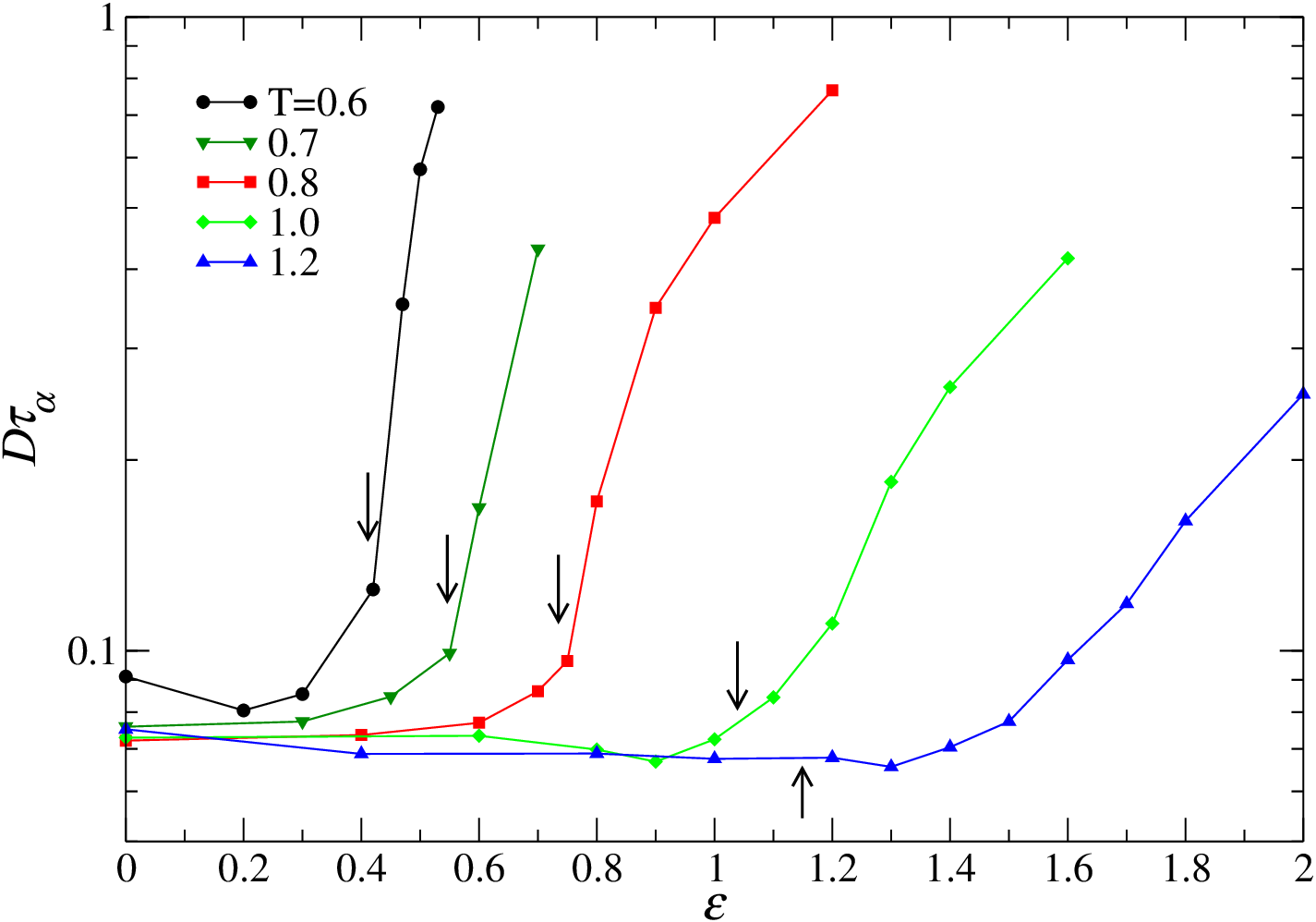}
\caption{\label{fig:se_se} 
Decoupling of the diffusion constant and the structural 
relaxation is established since the product $D \tau_\alpha$ 
increases dramatically with $\epsi$. The arrows mark the crossover
$\epsi^*(T)$.} 
\end{figure}

An inspection of Fig.~\ref{fig:tau_D} suggests that diffusion and 
structural relaxation both slow down dramatically in the high-$Q$
regime, but their behaviour is also strongly decoupled, 
as $D^{-1}$ appears to change less than $\tau_\alpha$. 
This is demonstrated more clearly in Fig.~\ref{fig:se_se}, where we show 
the product $D \tau_\alpha$ for the $A$ particles. We observe that 
$D \tau_\alpha$ remains constant until $\varepsilon$ 
reaches $\epsi^*$, above which it starts to grow significantly.
Interestingly, the deviation is visible for both species, but it is 
less pronounced for the small particles (not shown).  

Such violation of the Stokes-Einstein relation, $D \tau_\alpha 
\approx D \eta \sim cst$, 
is expected in systems whose dynamics is highly heterogeneous, such as 
supercooled liquids~\cite{ediger00}, although other sources 
of decoupling have been identified in simple fluids \cite{zamponi2}.
The coexistence of slow and fast particles in the sample leads to a 
decoupling between $\tau_\alpha$ and $D$, because the former is mostly 
dominated by the slow particles, whereas the latter is dominated 
by the fast ones. We remark, however, that the Stokes-Einstein 
violation at large $\varepsilon$ is more 
pronounced than the one observed by decreasing temperature in the 
unconstrained model.
This suggests that the physical origin of this decoupling might be
somewhat different in the constrained model than in the bulk.

A first hypothesis to explain the decoupling 
is the existence of large-scale spatial correlations in the 
dynamics~\cite{tarjus95}. 
This hypothesis is reasonable, since static correlations 
are promoted by the coupling field $\epsi$, as testified by the 
maximum in the static susceptibility $\chi$ in Fig.~\ref{fig:qave}.

\begin{figure}\centering
\includegraphics[width=8cm,clip]{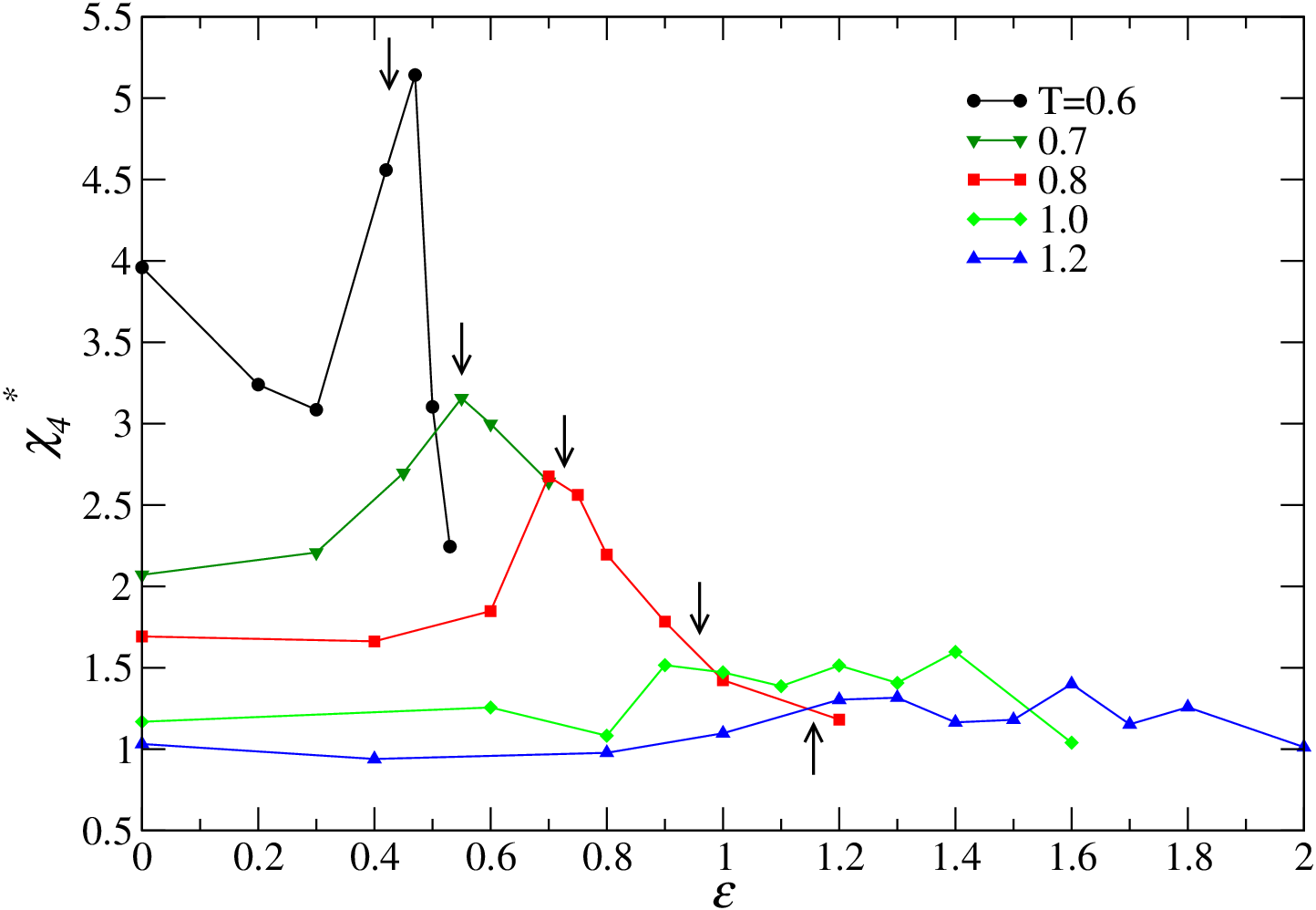}
\caption{\label{fig:se_chi4}  Evolution of the 
maximum of the four-point dynamic susceptibility $\chi_4^*$ 
as a function of $\varepsilon$. Spatial correlations 
are maximum near the static crossover $\epsi^*(T)$
(indicated by arrows).}
\end{figure}

To quantify spatially heterogeneous dynamics in a more precise manner, 
we evaluate the four-point dynamic 
susceptibility~\cite{toninelli,berthier_spontaneous_2007,berthier_spontaneous_2007-1},
\begin{equation} 
\chi_4(t)=N \left[ \langle 
f_s(k,t)^2\rangle - \langle f_s(k,t)\rangle^2 \right], 
\end{equation}
where $f_s(k,t)= \frac{1}{N} \sum_i 
\exp [ i {\bf k} \cdot ({\bf r}_i(t) - {\bf r}_i(0)) ]$
represents the instantaneous value of the self-intermediate 
scattering function.
In supercooled liquids, $\chi_4(t)$ is characterized by a single peak, 
located around $\tau_\alpha$, whose height $\chi_4^*$ provides a simple measure 
of the degree of heterogeneity of the dynamics (it scales 
roughly as the dynamic correlation volume). We find that 
the constrained system displays similar peaks in $\chi_4(t)$, 
and we represent the variation of $\chi_4^*$ as a 
function of $\varepsilon$ in Figure~\ref{fig:se_chi4} for 
all studied temperatures.

At high temperature, $\chi_4^*$ is essentially 
constant and practically independent of $\varepsilon$.
When $T < 1.0$ the behavior changes and $\chi_4^*$ varies non-monotonically 
as a function of $\varepsilon$, with a maximum located very close to 
the crossover coupling $\varepsilon^*$ defined from the maximum 
of the static overlap fluctuations. This shows that 
spatial correlations in the dynamics are essentially 
slaved to the spatial fluctuations of the static overlap.
In particular, we find that $\chi_4^*$ decreases rapidly 
as $\epsi$ is increased beyond $\epsi^*(T)$. This 
implies that the collective dynamics in the localized regime is actually 
strongly suppressed and that particles move instead in a spatially uncorrelated 
manner. Therefore, the strong decoupling reported in Fig.~\ref{fig:se_se} 
does not stem from the coexistence of dynamically correlated domains. 

Nevertheless, it is interesting to note 
that the behaviour of the static and dynamic susceptibilities 
$\chi$ and $\chi_4$ is similar only when $\epsi$ is large, 
but it differs dramatically when $\epsi \to 0$ where 
dynamic fluctuations increase rapidly as $T$ decreases whereas 
static ones are essentially independent of temperature for the 
unconstrained system at $\epsi = 0$. The fact that static and dynamic
fluctuations have distinct temperature dependences in the moderately 
supercooled regime was noted 
before~\cite{Kob-NatPhys2011,berthier_static_2012,Charbonneau_Charbonneau_Tarjus_2012}.

A final notable feature of our data is observed at the lowest studied
temperature at $T=0.6$ in Fig.~\ref{fig:se_chi4}. 
Despite the limited dynamic range we can 
access, the data reveal an additional feature, since  
$\chi_4^*$ is a decreasing function of $\varepsilon$ at small $\varepsilon$.
Such a decrease of $\chi_4^*$ is reminiscent of the one observed 
with randomly pinned particles~\cite{jack2014,kob_nonlinear_2014}.
For such systems, the relevant order parameter is the concentration of 
pinned particles, which plays an analogous role to the coupling 
$\varepsilon$ in the present model.
In the randomly pinned systems, however, there was no trace of a subsequent 
increase of $\chi_4^*$ as a function of the concentration of pinned 
particles \cite{jack2014}, presumably because the crossover line cannot 
be easily approached in the time window available to conventional computer 
simulations in the case of random pinning.

\begin{figure}\centering
\includegraphics[width=8cm,clip]{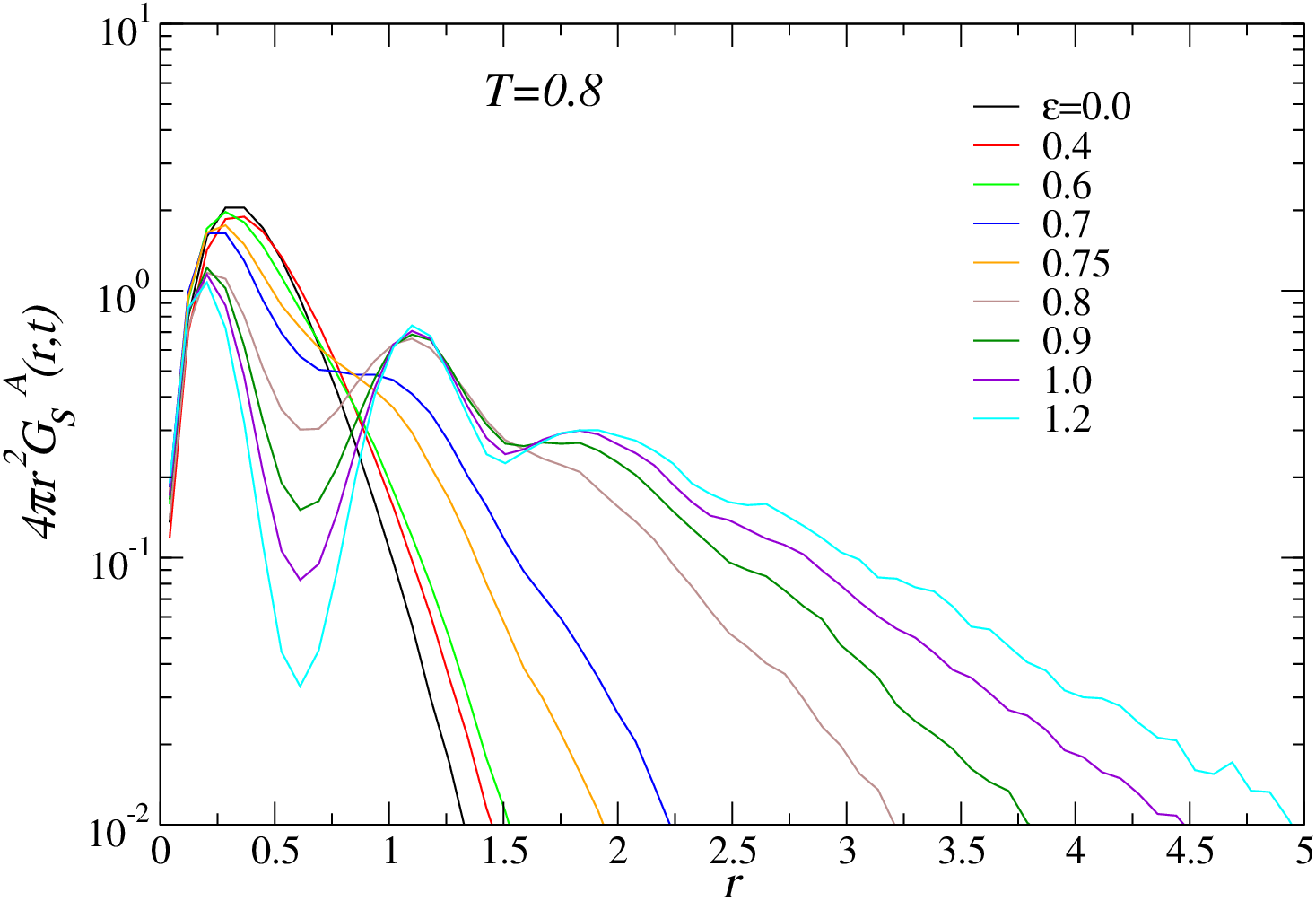}
\caption{\label{fig:vanhove} Self-part of the van-Hove correlation function
$4\pi r^2 G_s(r,t)$ for large particles for several $\varepsilon$ 
at $T=0.8$. The distributions are evaluated at times $t$ such that 
$F_s(k,t)=0.2$. They broaden considerably as $\epsi$ increases,
and the tails develop a multi-peak structure resulting from 
discrete particle hopping.} 
\end{figure}

The analysis of the four-point dynamic susceptibility shows that 
the particle dynamics becomes essentially uncorrelated 
in the high-overlap regime. Therefore, the explanation of 
the strong decoupling behavior between diffusion and 
alpha-relaxation time should be sought in the properties
of single particle motion. To elucidate this aspect, 
we evalute the self-part of the van-Hove correlation function,
\begin{equation}
G_s(r,t) = \left\langle 
\sum_{i} \delta (r - |{\bf r}_i(t) - {\bf r}_i(0) |) \right\rangle.
\end{equation}
In Fig.~\ref{fig:vanhove}, we show the evolution of the distribution
of single particle displacements for increasing $\epsi$, 
adjusting for each value of $\epsi$ the timescale such that 
$F_s(k,t) \approx 0.2$ (this allows us to probe displacements that are 
large enough to better reveal the structure of the van-Hove function). 
The distributions of displacements are essentially featureless 
and close to a Gaussian for small $\epsi$, but they broaden 
considerably when $\epsi > \epsi^*(T)$. 
This considerable broadening suggests that at any given time, 
the system is composed of fast-moving and slow-moving particles
that coexist in space. This feature alone is sufficient to 
account quantitatively for a strong decoupling phenomenon \cite{bcg05}.
It is interesting to note that the coexistence in space of fast and slow
particles with increasingly heterogeneous dynamics 
is not associated to growing spatial correlations of the dynamic
relaxation; this is instead a purely local phenomenon.     

The data in Fig.~\ref{fig:vanhove} not only show that the 
distributions broaden with $\epsi$, since the shape of the distributions
is also changing qualitatively. The distributions are strongly 
non-Gaussian at large $\epsi$, and the emergence of secondary
peaks is also obvious. These peaks suggest that the particle dynamics
evolves from a continuous diffusive process at small $\epsi$,
to a slow, intermittent, hopping process at large $\epsi$. 
Physically, this means that particles jump mostly from one site to
another, where the `sites' are defined by the positions of the particles 
in the reference configuration. This interpretation is harmonious
with the observation of the formation of dimers discussed
in Sec.~\ref{sec:statics}. Finally, 
we note that particles of both species are involved in the hops,
although the behaviour seems less pronounced for small particles,
whose van-Hove functions are less structured than for large particles 
(data not shown).  This observation
contrasts with the behavior of the unconstrained model at low
temperatures around $T_\textrm{mct}$, where the jump dynamics 
mostly involve the smaller $B$ particles.

\section{Conclusions}
\label{sec:conclusion}

We have used conventional Monte-Carlo simulations to 
study a viscous liquid coupled to a frozen reference configuration 
of the same liquid at the same temperature.
Our exploration of the high-temperature portion of the $(\epsi, T)$ 
phase diagram reveals a `Widom line' of thermodynamic and dynamic anomalies 
signalling the crossover from an uncorrelated liquid regime to a localized one, 
in which particles of the system are strongly constrained to reside near 
the sites of the reference configuration.
These results as compatible with the existence of a line of first-order 
transitions at lower temperature terminating at a second-order critical point.
Our results show that computational approaches to this problem
suffer from two distinct sources of dynamic slowing down. 
First, as in any phase transition, the order parameter (here the 
global overlap) displays large fluctuations near coexistence, reflecting
the emergence of a non-trivial free-energy profile. Second, 
the single-particle motion becomes very slow in the high-overlap 
region because particles must hop on the sites of the reference 
configuration in a highly
constrained manner in order for the global overlap to remain large.
Altogether, this means that at least two distinct strategies 
need to be implemented in order to study the phase diagram
at lower temperatures to overcome both the free-energy barriers 
near the phase transition and the slowing down of the particles motion in the 
high-overlap regime. Work is in progress in this direction~\cite{berthier2015}, 
which will hopefully allow for a direct study of the 
equilibrium phase diagram of coupled viscous liquids. 

\section*{Acknowledgements}

We are very pleased and honoured to contribute an article to this Special Issue 
in honour of J.-P. Hansen whose work, career and integrity set an 
admirable example to all of us. 
The research leading to these results has received funding
from the European Research Council under the European Union's Seventh
Framework Programme (FP7/2007-2013) / ERC Grant agreement No 306845.

\bibliographystyle{tMPH}
\bibliography{paper}

\end{document}